May 1995

# Large Logarithm Behaviour of $e^+e^-$ Jet Cross Sections and Event Shape Distributions in $O(\alpha_s^2)$

G. Kramer[a], H. Spiesberger[b]
[a]II. Institut für Theoretische Physik*,†
Universität Hamburg
D - 22761 Hamburg, Germany
[b]Fakultät für Physik†
Universität Bielefeld
D - 33615 Bielefeld, Germany

**Abstract**

We have calculated the leading and next-to-leading logarithm coefficients of $O(\alpha_s^2)$ $e^+e^-$ annihilation jet cross sections, thrust distribution and energy-energy correlation in the two-jet limit when the jet resolution and the event shape variables vanish. We have compared our results with expectations based on leading logarithm approximations used to resum the pertubative cross sections where this is possible. There is good agreement for the leading and next-to-leading coefficients of jet cross sections in the Durham scheme. Also for the thrust distribution and energy-energy correlation we find results which are consistent with the leading logarithm predictions.

* Supported by Bundesministerium für Forschung und Technologie, Bonn, Germany, Contract 05 6HH93P(5).

† Supported by the EEC Program 'Human Capital and Mobility' through Network 'Physics at High Energy Colliders' under Contract CHRX-CT93-0357 (DG12 COMA).

# 1 Introduction

The four LEP experiments have given us very accurate data for multijet cross sections and event shape distributions in $e^+e^-$ annihilation at the $Z$ resonance. These data have been successfully compared to perturbative $QCD$ and presently constitute one of the best ways to determine the $QCD$ coupling constant $\alpha_s$. Observables related to the structure of hadronic final states produced in $e^+e^-$ annihilation, such as multijet rates, thrust, or energy-energy correlation have been calculated exactly up to second order in $QCD$ perturbation theory [1, 2]. These calculations have been widely used by experimental groups at the PETRA, PEP, TRISTAN, SLC and LEP colliders for determining the strong coupling constant $\alpha_s$ [3, 4].

The experimental accuracy of the data from these experiments, in particular those coming from LEP, is now very high so that the $\alpha_s$ determinations are limited mainly by the theoretical uncertainties connected with using second order ($O(\alpha_s^2)$) perturbation theory.

Under normal circumstances we might expect that the third order term and beyond is $\alpha_s^3$ times a number of order unity which would yield a few percent corrections of the leading $\alpha_s$ term at LEP energies. This expectation is only reasonable, however, when the jet resolution parameters or the event shape variables are far from the two-jet limit. It is well known that for a dimensionless jet resolution or event shape variable $y$ vanishing in the two-jet limit, the leading behaviour of the distribution in the n'th order and in the limit $y \to 0$ is of the following form

$$\frac{1}{\sigma}\,\frac{d\sigma}{dy} \sim \alpha_s^n \frac{1}{y}\left(\ln\frac{1}{y}\right)^{2n-1} \tag{1}$$

so that the jet rate or the normalized event shape cross section $R(y)$ defined by

$$R(y) = \int_0^y \frac{1}{\sigma}\frac{d\sigma}{dy'}dy' \tag{2}$$

has the behaviour (considering that at $y = 0$ real and virtual singularities are cancelled)

$$R(y) = \sum_n \alpha_s^n R_n(y), \tag{3}$$

$$R_n(y) \simeq L^{2n} \tag{4}$$

where $L = \ln y$. It is clear that for small $y$ approaching the two-jet limit the perturbative result is not reliable anymore since $\alpha_s L^2$ is not small at high energies even when $\alpha_s$ is already small enough. For such a situation the perturbative series must be summed in order to obtain a more reliable estimate for the considered observable. A general summation is out of reach. However, for a selection of cluster and shape variables and for the logarithmically dominant terms in such cases where these leading logarithmic contributions exponentiate, this seems to be possible [5, 6, 7, 8, 9, 10, 11, 12]. These resummed expressions are valid only in leading and next-to-leading logarithmic accuracy, i.e. only in the limit $y \to 0$. Their use in the data analysis is doubtful, since $y$ values below 0.002 can not be reached and as far as we know, this value is not small enough to allow neglecting subleading terms.

In order to obtain reliable predictions also for larger values of $y$, one combines the resummed and fixed-order calculations. This can be done in several ways [8]. The most popular matching schemes start with the requirement that the leading (LL) and next-to-leading logarithmic (NLL) terms in the resummed expressions if expanded in powers of $\alpha_s$ up to $O(\alpha_s^2)$ should agree with the LL and NLL terms of the fixed-order results. It is easy to calculate the $O(\alpha_s)$ LL and NLL coefficients and compare them to the $O(\alpha_s)$ coefficient in the resummed prediction. The $O(\alpha_s^2)$ coefficients, however, can be obtained only from extensive numerical evaluations. In some cases it has been checked explicitly that the LL and NLL coefficients up to $O(\alpha_s^2)$ in the resummed results agree with those obtained by fitting the asymptotic behaviour of the fixed order (up to $O(\alpha_s^2)$) cross sections. Such ascertainments are made for the multijet cross sections defined with the $k_T$ clustering algorithm [7], for the distribution of the jet broadening [10] and the heavy jet mass [9], for the thrust distribution [8] and for the back-to-back energy-energy correlation [12]. For those cases where the results of such computations are given explicitly in terms of actual numbers (for thrust, heavy jet mass, jet broadening and energy-energy correlation) one sees that these $O(\alpha_s^2)$ LL and NLL coefficients have appreciable errors (sometimes of the order of 100 % for the LL coefficient). Therefore these results show at best that the $O(\alpha_s^2)$ LL and NLL coefficients are consistent within the large errors with the LL and NLL $O(\alpha_s^2)$ terms in the resummed predictions. From these checks it is not excluded that even the leading logarithm coefficients in the complete $O(\alpha_s^2)$ calculations differ from the corresponding terms in the 'all-order results'. If this would be true the resummed cross sections and shape variable distributions would be much less useful and the results for $\alpha_s$ obtained with the 'all-order' formulas would be in doubt.

In this work we make a new effort to calculate the LL and NLL coefficients of the complete $O(\alpha_s^2)$ theory for $e^+e^-$ annihilation into 2, 3 and 4 jets. We base our calculations on the work of B. Lampe and one of us [1]. In this work multijet cross sections and the thrust distribution have been calculated by a combination of analytical and numerical methods using the so-called phase space slicing method. This differs from the so-called subtraction method used in the Monte Carlo integration program EVENT [2] based on the matrix elements [13]. Of course, we do not expect that these two methods lead to different answers. We hope that with our method we shall get numerical results with smaller errors for the LL and NLL coefficients. These coefficients will also be calculated for the multijet cross sections with invariant mass clustering which do not exponentiate but for which analytical predictions for the LL and NLL coefficients in $O(\alpha_s^2)$ have been derived [5] using leading logarithm approximation methods for the evaluation of the tree graph matrix elements.

The plan of the paper is as follows. In sect. 2 we recapitulate the method used to calculate the various multijet cross sections and event shape distributions in the earlier work [1]. Then we explain how these methods can be applied for the phase space slicing method. In sect. 3 we present the results and discuss them in view of the expectations from the resummed calculations.

# 2 Calculational Methods

The calculation of the event shape distributions and the multijet cross sections proceeds on the basis of the approach described in detail in [1]. It starts from known $O(\alpha_s^2)$ matrix elements for $e^+e^- \to 3$ *partons* [13, 14] and $e^+e^- \to 4$ *partons* [15]. For obtaining finite results in which all infrared and collinear singularities have been cancelled one introduces a resolution cut to separate the 4-parton phase space into a genuine 4-parton final state region and a region containing states in which two of the partons are combined into one jet. For this recombination we demand for two partons with labels $i$ and $j$ (the 4-momenta of the partons are denoted by $p_i$, $i = 1, 2, 3, 4$)

$$y_{ij} = (p_i + p_j)^2/q^2 < y \tag{5}$$

where $y$ is the value of the resolution cut and $\sqrt{q^2}$ is the total center-of-mass energy. This invariant mass resolution is most convenient to perform analytical calculations. The integration over the degenerate states inside the 4-parton phase space is performed using a partial fractioning decomposition of the 4-parton matrix elements which allows a separation of the singular and non-singular contributions. Then the calculation of the 3-jet cross section consists of three parts:

(i) The first part contains the singular parts of the 4-parton final state (real contribution) integrated over the infrared/collinear singular region inside the resolution cut together with the virtual corrections to the 3-parton final state.

(ii) The second part contains the remaining non-singular contributions of the 4-parton final state integrated over the same regions as in (i).

(iii) The third part contains all contributions outside the singular region of (i) where two partons are again combined into one jet.

The 4-jet cross section in $O(\alpha_s^2)$ is obtained from the phase space region containing genuine 4-parton final states, i.e. where $y_{ij} \geq y$ for all $i$, $j$.

The calculation of the real contributions of (i) proceeds as follows. The cross section for $e^+e^- \to q\bar{q}gg$ (the $q\bar{q}q\bar{q}$ final state which is less singular is treated similarly) has the general form

$$d^5\sigma = \left(\frac{\alpha_s}{2\pi}\right)^2 f(y_{ij}) dPS^{(4)}. \tag{6}$$

The right-hand side of (6) contains pole terms proportional to $y_{ij}^{-1}$ ($ij$ = 13, 14, 23, 24, 34) which are separated from each other by partial fractioning (see [1] for details). For example the contribution proportional to the colour factor $C_F^2$ has the structure

$$f(y_{ij}) = \frac{A_{13}}{y_{13}} + (1 \leftrightarrow 2) + (3 \leftrightarrow 4) + (1 \leftrightarrow 2,\ 3 \leftrightarrow 4) \tag{7}$$

where the label of momenta is $1, 2, 3, 4 = q, \bar{q}, g, g$. The terms proportional to $y_{13}^{-1}$ and $y_{14}^{-1}$, respectively $y_{23}^{-1}$ and $y_{24}^{-1}$, contain the singularities where the gluon is infrared and/or collinear with the quark, respectively the antiquark. They produce the dominant contributions to the $O(\alpha_s^2)$ cross section for $e^+e^- \to 3\ jets$ and must be integrated over the unresolved regions $y_{13} < y$, $y_{23} < y$ etc., where $qg$, and $\bar{q}g$ are recombined into one jet. This integration is performed only for one of the four terms in (7) which are related to each other by a permutation of the momentum labels. The four terms in (7) are separated into so-called singular and non-singular terms. The singular terms are regularized by dimensional regularization. The singularities after integration compensate against the singularities in the $O(\alpha_s^2)$ one-loop corrections to $e^+e^- \to q\bar{q}g$. The integration is done analytically to isolate the remaining finite terms. The result is the contribution (i) referred to above.

The other two contributions (ii) and (iii) come exclusively from the real diagrams for $e^+e^- \to q\bar{q}gg$. The second part (ii) consists just of the non-singular terms in $A_{13}/y_{13}$ etc.

in (7) which are integrated over the unresolved region $y_{13} < y$ etc. They can be obtained by numerical integration in four dimensions and do not take part in the cancellation of the infrared/collinear singularities between real and virtual contributions. The third part (iii) is concerned with the resolved $qg$ region $y_{13} > y$ (in the first term of (7)). In this region those $qg$, $\bar{q}g$, $gg$ and $q\bar{q}$ recombinations must be done which are not singular, i.e. one has to integrate over the regions $y_{23} < y, y_{14} < y$ etc. which contain just three jets. For this integration the boundaries of the unresolved $qg$, $\bar{q}g$ etc. regions are very complicated so that Monte Carlo integration methods must be applied.

A complication arises when we want to define 3-jet variables for those contributions where two of the four final state partons have been recombined into one jet. There is no unique way to form 3-jet variables out of the 4-parton variables of the $q\bar{q}gg$ final state. In [1] two schemes have been considered: (1) the so-called KL scheme where $y_{134}$ ($y_{ijk} = (p_i + p_j + p_k)^2/s$) and $y_{24}$ are defined as 3-jet variables in the case where parton 1 and parton 3 are recombined. According to the 4-parton kinematics the third variable is $y_{123} - y_{13}$ (equivalently for the $1 \leftrightarrow 2,\ 3 \leftrightarrow 4$ and $1 \leftrightarrow 2,\ 3 \leftrightarrow 4$ permutations); (2) the so-called KL' scheme where the 3-jet variables are $y_{134}$ and $y_{123}$ together with $y_{24} - y_{13}$ according to the 4-parton constraint. Of course these are not the only possibilities. This non-uniqueness of the 3-jet variables in $O(\alpha_s^2)$ is unavoidable when one cancels the infrared/collinear divergences between the virtual and real contributions which is only possible when the $O(\alpha_s)$ cross section is factored out in the singular 4-parton contributions. The differences in the 3-jet variables disappear in the limit $y_{13} \to 0$ (equivalently for the other permutations). Thus, when the cut parameter $y$ is chosen very small we expect that the difference between the two schemes KL and KL' disappears.

By introducing the invariant mass cut in the partial fractioning decomposition we can use our approach in two ways: (a) By choosing the cut value $y$ extremely small, the cut just serves the purpose to cancel the infrared/collinear divergences and any dependence on it drops out when we add the appropriate '3-jet' and '4-jet' contributions. Then we can calculate any distribution we are interested in. This works even in cases where for the chosen jet variable the infrared/collinear cancellation cannot be obtained through an analytic calculation. In addition, this approach has the advantage that contribution (ii), which is the most difficult to calculate, cannot contribute to the asymptotic logarithmic behaviour. In this case we will use $y_0$ to denote the cut value (slicing parameter), whereas we will still use $y$ for the jet resolution parameter (see Sec. 3 for details). (b) For finite $y$, i.e. identifying the jet resolution $y$ with the slicing parameter $y_0$, we can calculate the dependence of the multijet rates on $y$ for the special case of invariant mass recombination as defined in the KL' scheme.

When presenting our results we shall distinguish the following contributions for which we shall list the results separately in all the cases where this is possible and appropriate. First we consider the contribution of the singular region (i) which contains the virtual part and those parts of the 4-parton region needed to cancel the infrared/collinear divergences, i.e. just the singular parts. In some cases we separate terms which originate from the virtual corrections and which do not take part in the infrared/collinear cancellation. Second we consider the contribution from the non-singular 4-parton terms which correspond to the region (ii) above. Third we list the contributions from the region (iii) which comes from the region where the invariant in the pole term in (6) is above the $y$ cut. This region includes also the genuine 4-jet contribution for all cases of shape variable distributions.

We write every normalized event shape cross section or multijet rate $R(y)$ defined in (2) as an expansion in $L$ in the following form

$$R(y) = \frac{\sigma_0}{\sigma}\left[R_0 + \frac{\alpha_s}{2\pi}\; R_1(y) + \left(\frac{\alpha_s}{2\pi}\right)^2\; R_2(y)\right] \tag{8}$$

where

$$R_n(y) = \sum_{m=0}^{2n} R_{nm}\; L^m + D_n(y). \tag{9}$$

The remainder functions $D_n(y)$ vanish in the limit $y \to 0$. We are interested only in $R_2(y)$, since $R_1(y)$ is well known. The total cross section $\sigma$ is known up to $O(\alpha_s^3)$ and has been factored out. Therefore the coefficient functions $R_n$ are normalized with the zeroth order cross section $\sigma_0$ instead of the full $\sigma$ as in (2) which is usually employed. $R_2(y)$ is decomposed according to the different colour factors in the following form

$$R_2(y) = C_F^2 R_2^C(y) + C_F N_C R_2^N(y) + C_F N_f R_2^T(y) \tag{10}$$

where $N_f$ is the number of flavours, $N_C$ the number of colours and $C_F = (N_C^2 - 1)/2N_C$ as usual.

The formulas for the calculation of the contributions (i), (ii), and (iii) to the differential cross sections for $q\bar{q}g$ and $q\bar{q}gg$ final states have been implemented into a Monte Carlo integration program allowing us to calculate jet rates and event-shape distributions with almost arbitrary recombination prescriptions and phase-space limits. The implementation takes advantage of the partial fractioning of the cross section into terms exhibiting the leading singular behaviour in the form of factors $1/y_{ij}$. The simple mapping $y_{ij} \to \ln y_{ij}$ renders the integrand into a flat function of the new integration variable, thus improving the reachable accuracy of the Monte Carlo integration considerably.

Each contribution for each value of the jet resolution parameter $y$ was determined in a separate Monte Carlo run, each time using between $10^5$ and $10^7$ points depending on the behaviour of the integrand. The resulting relative Monte Carlo error was typically of the order of $10^{-3}$ or smaller. Since the final results for the jet rates are obtained from summing at least three contributions which come with different signs, the jet rates themselves, however, suffer from large cancellations so that their relative errors are only at the per cent level. This is one of the reasons why the coefficients of $\ln y$ for separate contributions could be obtained with smaller errors than for the complete jet rates.

To obtain the $y$-dependence of jet rates, $R_2(y)$ has been calculated for $y$-values chosen below $L = \ln y = -5$ (i.e. $y \simeq 7 \cdot 10^{-3}$). This way large $y$-values where non-logarithmic contributions (e.g. terms of the type $y \ln^n y$) are probably important can be excluded and reasonable fits taking into account only powers of $L$ can be expected. Data points have been chosen with equidistantly distributed $L$ with $\Delta L = 0.2$. In our first set of results (case (a)) where a fixed value of the slicing parameter $y_0 = 10^{-5}$ was chosen, the data points extended down to $L = -10$; thus 26 data points are available for the fit in this case. The second set of results with $y = y_0$ (case (b)) comprises data points down to $L = -14$ (i.e. $y \simeq 8 \cdot 10^{-7}$), thus in this case we have 46 data points for the fit.

The Monte Carlo results for $R_2(y)$ have been fitted to the functional dependence (9) with $D_n(y) = 0$ using the program MINUIT [16]. The fitted values for the coefficients $R_{nm}$ are accepted and displayed in the tables if MINUIT obtained a reasonably small $\chi^2$, $\chi^2$/d.o.f. $\simeq O(1)$. Also, the quality of the fit was tested by varying the range of $L$ used in the fit and choosing subsets of the data points for $R_2(y)$. Usually a reasonable $\chi^2$ could be obtained when fitting the whole range of Monte Carlo data, i.e. for $-5 \geq L \geq -10$ for complete jet rates and $-5 \geq L \geq -14$ for their separate contributions.

For the $C_F N_C$ contributions to the 3–jet rates in the JADE-E and E0 schemes, however, no acceptable fit could be obtained when including the smallest $y$ values. The reason for this is a logarithmic, though integrable, divergence of the integrand for the contribution (ii), rendering the Monte Carlo integration unstable for very small $y$ values. Thus the range of data points used for the fit had to be reduced to results with $L > -9$. A further reduction of the $L$ range had not much influence on the fitted coefficients; however, due to the reduced number of data points, their errors are quite large.

The event-shape distributions (thrust and energy-energy correlation) have been calculated in bins of the shape variable ($L_\tau = \ln\tau = \ln(1-T)$ for thrust, $L_\eta = \ln(1/\eta) = \ln(2/(1+\cos\chi))$ for the energy-energy correlation) again with equidistant bins of width $\Delta L = 0.2$. The range of values for $|L|$ used in the fits was $5 \leq |L| \leq 14$. In both cases, the slicing parameter $y_0$ was chosen large, $y_0 = 0.1$, in order to reduce the contribution from region (iii). This contribution comes from $q\bar{q}gg$ events and for the calculation of the event shape variables one would have to use the 4-particle kinematics, whereas for contributions (i) and (ii) 3-particle kinematics determines the event-shape variables. By choosing a large value for $y_0$, the contribution from region (iii) is vanishing in the asymptotic region of large (negative) values for the shape variables $L_\tau$ and $L_\eta$. Thus this part cannot contribute to the coefficients of the asymptotic expansion (9) and has been neglected in the fits.

As was the case for the $C_F N_C$ contribution to the 3-jet rates, also the $C_F N_C$ part of the thrust distribution suffers from numerical instabilities for large negative $L$. Here again, the range of $L$ values that could be used for the fits had to be reduced to values $|L| < 9$ for the contribution of region (ii).

# 3 Results

We start with the multijet rates calculated in different recombination prescriptions. We consider the JADE-E0, -E and the Durham-E0 scheme. First we introduce the slicing parameter $y_0$. The slicing of the phase space is defined with the invariant mass

$$m_{kl}^2 = (p_k + p_l)^2 = 2p_k p_l, \tag{11}$$

$$y_{kl} = m_{kl}^2/q^2. \tag{12}$$

For $y_{kl} < y_0$ the 4-parton contributions are integrated analytically and combined with the virtual corrections. This is the contribution (i) described in sect. 2. For $y_{kl} > y_0$ we obtain the contribution of the real diagrams for $e^+e^- \to q\bar{q}gg$. In the following we consider only $R_2^C$ and $R_2^N$. $R_2^T$ which comes from $e^+e^- \to q\bar{q}q\bar{q}$ and the corresponding virtual terms is less singular and will be discussed at the end. We choose $y_0 = 10^{-5}$ which is small enough, so that the contribution in region (ii), i.e. the non-singular parts make a negligible contribution. This has been checked explicitly. In the region $y_{kl} > y_0$ the jet cross sections are calculated from the 4-parton final state $q(p_1)+\bar{q}(p_2)+g(p_3)+g(p_4)$ with the well-known JADE [17] algorithm which consists of the following iterative procedure ("jet algorithm"):
1) We define a resolution parameter $y$.
2) For every pair $p_k$, $p_l$ of parton momenta in a 4-parton event we compute the "distance measure" $M_{kl}^2$ (which can, but need not, be identical to the invariant mass $m_{kl}^2$) and define $y_{kl} = M_{kl}^2/q^2$ $(q = p_1 + p_2 + p_3 + p_4)$.
3) If $y_{ij}$ is the smallest value of all $y_{kl}$ computed in 2) and $y_{ij} < y$ we combine the partons with momenta $p_i$ and $p_j$ into a single jet with momentum $p_{ij}$ according to a recombination procedure.
4) This procedure is repeated until all pairs of objects (i.e. partons and/or jets) have $y_{kl} > y$.
The remaining objects are identified with jets. Events with 2, 3, or 4 jets contribute to the 2-, 3-, or 4-jet cross sections, respectively. For the recombination we consider only the E0 and E schemes. In the E0 scheme, $M_{kl}^2 = 2\ E_k\ E_l\ (1 - \cos\theta_{kl})$ is the invariant mass squared for massless partons. The recombined 4-momentum $p_{ij} = (E_{ij}, \vec{p_{ij}})$ is defined by

$$E_{ij} = E_i + E_j, \quad \vec{p_{ij}} = \frac{E_{ij}}{|\vec{p_i} + \vec{p_j}|}(\vec{p_i} + \vec{p_j}). \tag{13}$$

This definition of $\vec{p_{ij}}$ has the effect that the combined jet is again massless so that in the recombination of $\vec{p_{ij}}$ with another parton the E0 scheme definition of $M_{kl}^2$ can be applied consistently. In the so-called E-scheme one uses $M_{kl}^2 = (p_k + p_l)^2$ and instead of (13) the recombination prescription $p_{ij} = p_i + p_j$ which is fully Lorentz invariant. We remark that for the slicing of the phase space with $y_0$, the slicing condition $y_{kl} < y_0$ is used only once to recombine two partons of the 4-parton final state into one jet. This serves to separate singular contributions in the 4-parton final states and isolates 3-jet final states where the resulting 3-jet variables are always defined in the KL' scheme. The slicing parameter $y_0$ is not used in the jet algorithm.

The Durham, or $k_T$-, algorithm [7] is defined in a similar way with the iterative procedure 1) - 4) and the "distance measure" $M_{kl}$ equal to the transverse momentum as jet resolution variable, i.e. the test variable in step 2) is

$$y_{kl} = 2\min(E_k^2, E_l^2)\ (1 - \cos\theta_{kl})/q^2. \tag{14}$$

For the recombination, i.e. for the calculation of $p_{ij}$, one can use again several schemes (E0, E etc.) as in the JADE algorithm. We shall limit ourselves to the E0 recombination scheme in this case.

Our results are presented in Table 1 for the JADE-E0 algorithm, in Table 2 for the JADE-E algorithm and in Table 3 for the Durham-E0 algorithm. $n$ denotes the number of jets and $R_{2m}$ was defined in (9). We distinguish between the $C_F^2$ and $C_F N_C$ contributions on the basis of the decomposition (10). The LL and NLL coefficients of the leading logarithm calculations as reported by Catani [5] to which we compare our results are as follows (for the $C_F^2$ ($C_F N_C$) coefficients, respectively): a) JADE-E0 and $n = 3$: $R_{24} = -11/3$ $(-1/6)$ and $R_{23} = -12$ $(-95/18)$, b) JADE-E0 and $n = 4$: $R_{24} = 3/2$ $(1/6)$ and $R_{23} = 6$ $(29/18)$, c) JADE-E $n = 3$ and $n = 4$: the same coefficients as in JADE-E0 with the only exception that for $n = 3$ the $C_F^2$ coefficient $R_{24}$ is $R_{24} = -19/6$, d) Durham-E0 and $n = 3$: $R_{24} = -1$ $(-1/12)$ and $R_{23} = -6$ $(-7/3)$ and

e) Durham-E0 and $n=4$: $R_{24}=1/2$ (1/12) and $R_{23}=3$ (10/9). For completeness we also give the results of our fit for $R_{22}$, $R_{21}$ and $R_{20}$ for which leading logarithm results are not available. These latter numbers are obtained with the constraint that $R_{24}$ and $R_{23}$ are equal to Catani's LL and NLL results. Also $R_{23}$ is calculated assuming the LL values for $R_{24}$.

The results for the Durham-E0 scheme presented in Tab. 3 show good agreement with the leading logarithm results. According to [5], they are identical to the results for the recombination schemes E, P and P0. The coefficients $R_{24}$ and $R_{23}$ for $n=4$ agree perfectly within the given errors with the leading logarithm results [5, 7] $R_{24}=1/2$, $R_{23}=3$ ($C_F^2$) and $R_{24}=1/12$, $R_{23}=10/9$ ($C_F N_C$). For $n=3$ the coefficients $R_{24}$ and $R_{23}$ agree also within errors with $R_{24}=-1$, $R_{23}=-6$ ($C_F^2$) and $R_{24}=-1/12$, $R_{23}=-7/3$ ($C_F N_C$). Here the errors are larger than for $n=4$. In total we can conclude that the $n=3$ and $n=4$ fixed order multijet cross section evaluated on the basis of the procedure of [1] are in very good agreement with the leading logarithm results obtained in [5, 7].

For the other two algorithms, the JADE-E0 and E scheme, the results are less precise and the agreement with the expected values is worse. First there is no difference between the E0 and E recombination schemes in the 4-jet rate since no recombination is necessary in this case. Therefore, for $n=4$, the coefficients $R_{2m}$ are equal in the two schemes. The numerical results for $R_{24}$ and $R_{23}$ for the $n=4$ $C_F^2$ and $C_F N_C$ coefficients are consistent with the leading logarithm result of [5]. The $C_F^2$ coefficient $R_{24}$ was first calculated by Brown and Stirling [18] using the eikonal approximation. They discovered the characteristic deviation of $R_{24}$ from the expectation based on the infrared singular behaviour of the cross sections for $e^+e^- \rightarrow q\bar{q}gg$ which yields for the $C_F^2$ coefficient $R_{24}=2$. The reduction of $R_{24}=(2-1/2)$ by $\Delta R_{24}=-1/2$ has its origin in the fact that the contribution from the phase space region $(p_3+p_4)^2/q^2 > y$, i.e. the region where the invariant mass of the two gluons is above the cut, has an effect on the leading coefficient $R_{24}$, although no infrared or collinear singularities in $y_{34}$ are present in the cross section for $e^+e^- \rightarrow q\bar{q}gg$. This has consequences for the leading double logarithm contributions in $n=2$ and 3 and spoils the exponentiation of the multijet cross section even in the abelian limit $N_C \rightarrow 0$. Our numerical result for $R_{24}$ supports this result and shows that it is not an artifact of the eikonal approximation used in [18]. $R_{24}$ ($C_F N_C$) and also the two $R_{23}$ coefficients are not affected by this particular $y_{34}$ contribution. They agree with the leading logarithm result of [5] within the numerical errors.

For the 3-jet rates the E0 and E scheme results differ. In the E0 scheme we get $R_{24}(C_F^2) = -1.63 \pm 0.69$ which is not in agreement with the LL result $R_{24}(C_F^2) = (-4+1/3)$ [5] (the term $+1/3$ in parentheses is the contribution from the $gg$ recombination mismatch according to [18]). In the E scheme $R_{24}(C_F^2) = -2.79 \pm 0.27$ is consistent with $R_{24}(C_F^2) = (-4+5/6)$ within the large error. This is also the case for $R_{24}(C_F N_C) = -1.6 \pm 2.0$ which is consistent with $R_{24}$ $(C_F N_C) = -1/6$. In the E0 scheme $R_{24}(C_F N_C) = -2.6 \pm 1.9$ is larger in absolute value compared to the leading double logarithmic result of [5] which is again $-1/6$. Since we have already deviations in the leading double logarithmic coefficient we can not expect that the next-to-leading coefficients agree better. Here even the signs of $R_{23}$ differ from the leading logarithm expectations. Whether these somewhat discouraging results are caused by our method of calculation, as for example the range of the fit intervals in $L=\ln y$, is not clear. Since the slicing parameter is only $y_0 = 10^{-5}$ the range of $y$ values which can be used to fit the asymptotic behaviour is limited to $y > y_0$ which perhaps is not sufficient to suppress non-logarithmic terms. Results for $R_{22}$, $R_{21}$, and $R_{20}$ are not given for the 3–jet rates in the E and E0 schemes since the errors given by the fit program are very large. The errors could have been reduced if a fit with fixed values for $R_{24}$ and $R_{23}$ was performed. Since our results for these LL and NLL coefficients do not agree well with the expected values from the leading logarithmic approximation, we do not attempt such a fit. Fixing $R_{24}$ and $R_{23}$ to the expected values would result in unacceptably large $\chi^2$. To investigate further the possible source of this disagreement, we have used a different method for calculating the coefficients $R_{2m}$ which will be described now.

We separate the contributions to $R_{24}$ and $R_{23}$ for the 3-jet cross section ($n = 3$) into the contribution from the singular terms, i.e. region (i) (see sect. 2), the non-singular terms, i.e. region (ii) and the contribution from the real 4-parton cross section, i.e. region (iii) which is the region where one $y_{ij} > y$ in the partial fractioning decomposition. The singular contribution (i) contains also the virtual correction terms which are not infrared and/or collinear singular. We have calculated these extra terms separately. In [13, 14] they are contained in the function $f(x_1, x_2)$ which was obtained there directly without any need for an infrared regularization. It is clear that this separation is not unique. The slicing parameter $y_0$ is identified with $y$ for the case of the JADE-E0 recombination scheme. The results presented in Tab. 4 are calculated in the various regions as a function of $y$ without any extra recombination added. The coefficients $R_{24}$ and $R_{23}$ are calculated by fitting the functional dependence on $y$ to (9) and (10). $y$ can be extended now to much lower values than with the pure slicing method done before.

In Tab. 4 we have obtained the following results for the JADE-E0 algorithm. Let us first discuss the $C_F^2$ part of $R_{24}$. It receives contributions from region (i) and region (iii). The contribution from (ii) is small and probably consistent with zero. There is also no contribution from the finite virtual term given by $f(x_1, x_2)$. The result in region (i) is consistent with $R_{24} = -4.0$, as expected for the singular contributions (see also [5]). The contribution from region (iii) is consistent with $R_{24} = 0.5$, so that the sum is $R_{24} = -3.5$. The obtained number with error is $R_{24} = -3.436 \pm 0.018$. This differs from Catani's LL value $R_{24} = -11/3 = (-4 + 1/3)$. The difference comes from region (iii). We think that this difference is real and outside the numerical error. Our results show definitely that $R_{24}$ is not given by the singular region (i) alone. There is a non-vanishing contribution from region (iii), which presumably is connected with the $gg$ recombination mismatch. We have not investigated this further.

The second column labelled $R_{24}$ in Tab. 4 gives the expected values in the different regions. The third and fourth column in Tab. 4 present the results of the simultaneous fit of $R_{24}$ and $R_{23}$ whereas the last column gives $R_{23}$ obtained with the hypothesis that $R_{24}$ is given as in the previous column. If we add the $C_F^2$ terms in the last column for $R_{23}$ we obtain $R_{23} = -12.098 \pm 0.041$ which is consistent with $-12.0$ expected from the leading logarithm result. From these numbers we can calculate the leading logarithm coefficients for the 2-jet cross section which is $R_{2m}$(2-jet)$= -(R_{2m}$(3-jet)$+R_{2m}$(4-jet)). Then we obtain for the 2-jet coefficients: $R_{24} = 1.976 \pm 0.015$ and $R_{23} = 5.76 \pm 0.20$. This agrees with $R_{24} = 2.0$ and $R_{23} = 6.0$ derived in an analytical calculation of the 2-jet cross section by B. Lampe and one of us [19].

Now we discuss the results for the $C_F N_C$ coefficients. The leading double logarithmic term receives contributions from all three regions (i), (ii) and (iii). It is interesting to note that also $f(x_1, x_2)$ contributes a term consistent with $-1/2$. The total sum for $R_{24}$ is $R_{24} = -1.905 \pm 0.011$. This is consistent with the value obtained in Tab. 1, but differs from the leading double logarithmic result of [5] which is $R_{24} = -1/6$. The reason for this difference is the contribution from the finite terms in (ii) which is large in absolute value. Then it is understandable that the next coefficient $R_{23}$, given in the fourth column, namely $R_{23} = -52.2 \pm 1.5$ (sum over all three regions), does not show any resemblance to the leading logarithm result $R_{23} = -95/18$. It is interesting to note that the leading logarithm expectations for $R_{24}$ and $R_{23}$ are consistent with the contribution from region (i). Therefore the problem might be that we obtain additional large terms from region (ii) and possibly from region (iii). In the case of $f(x_1, x_2)$ we give, in addition, results for the coefficients $R_{22}$ in parentheses. These latter results have been obtained assuming $R_{23} = 0$.

In an additional calculation we have investigated the separate contributions in the various regions (i), (ii), (iii) and $f(x_1, x_2)$ for the Durham-E0 scheme. The separation between the regions (i), (ii) on one side and (iii) on the other side is defined with the help of the invariant-mass cut. This means that the separation between the 3-jet contribution with one $y_{kl} < y$ and the genuine 4-parton region, where one $y_{kl} > y$, is done with the invariant-mass cut. The dependence on this invariant-mass cut must cancel in the sum of the three regions. Inside every region we have calculated the 3-jet fraction with the Durham-E0 boundaries and tabulated the results in Tab. 5. Since the singular terms, i.e. the integrand of the contribution in (i), is given analytically we achieve a much higher accuracy also in the sum. Note that the boundaries with the Durham variables are always inside the invariant-mass boundaries. We emphasize that the results in the different regions (i), (ii) and (iii) are not genuine 3-jet fractions in the Durham scheme, but their sum is. When we sum the contributions in the various regions we obtain for the $C_F^2$ coefficients $R_{24} = -0.984 \pm 0.020$, $R_{23} = -6.066 \pm 0.046$ in perfect agreement with the leading logarithm results $R_{24} = -1$, $R_{23} = -6$. In the same way we get for the sum of the $C_F N_C$ terms $R_{24} = -0.104 \pm 0.014$ and $R_{23} = -2.250 \pm 0.030$ which also agrees with the expected values $R_{24} = -1/12$ and $R_{23} = -7/3$. With the separation of the total contributions into the various regions where some of the integrations could be done analytically we achieved a much higher accuracy and this way we are able to verify the leading logarithm predictions with a much better accuracy than with our results in Tab. 3. Concerning the contribution of the function $f(x_1, x_2)$ which originates from the non-singular part of the virtual corrections we obtained $R_{24}(C_F N_C) = -0.0832 \pm 0.0009$ which is consistent with $R_{24}(C_F N_C) = -1/12$. This is a remarkable result in view of the fact that the leading logarithm calculations of [5, 7] are based on the summation of real contributions only and it is unclear how such a calculation could recover contributions from loop corrections. If we compare the result in region (i) with the contribution coming from $f(x_1, x_2)$ it seems that the contribution in (i) comes exclusively from the function $f(x_1, x_2)$.

Next we present our results for the two event shape distributions which we have investigated, namely the thrust distribution and the energy-energy correlation. Instead of the integrated distributions (2) we shall consider the differential distributions $dR/dy = 1/\sigma d\sigma/dy$. For the thrust distribution we have $y = \tau = (1 - T)$ and we write

$$\frac{1}{\sigma_0}\frac{d\sigma}{d\tau} = \frac{\alpha_s}{2\pi}\frac{dR_1}{d\tau} + \left(\frac{\alpha_s}{2\pi}\right)^2 \frac{dR_2}{d\tau}. \tag{15}$$

The contribution $\propto \alpha_s^2$ has the following expansion in $L_\tau = \ln\tau$, using (9),

$$\frac{dR_2}{d\tau} = \sum_{m=1}^{4} m R_{2m} \frac{1}{\tau} L_\tau^{m-1} + \frac{dD_2}{d\tau}. \tag{16}$$

First we have calculated the coefficients $(mR_{2m})$ for the two colour factors $C_F^2$ and $C_F N_C$. To separate the regions (i) and (ii) from (iii) we have introduced the slicing parameter $y_0 = 0.1$. This has the consequence that the thrust variable $T$ is always the 3-jet thrust since the contribution from region (iii) where we have to use the 4-jet thrust vanishes in the asymptotic region, i.e. for $|L_\tau| \to \infty$. The results for $mR_{2m}$ $(m = 1, 2, 3, 4)$ are presented in Tab. 6. We distinguish the contributions from the regions (i) and (ii), and the sum of (i) and (ii) which was fitted separately. Whereas the results for the separate contributions (i) and (ii) have been obtained from a fit without any assumption about the values of the LL coefficients, the coefficients $R_{23}$, $R_{22}$, and $R_{21}$ for the sum result from fits where $R_{24}$ was fixed at the corresponding leading logarithm expectation. We also give the results of a fit where the leading logarithm results for $R_{24}$ and $R_{23}$ are assumed and only $R_{22}$ and $R_{21}$ are determined from the fit. In those cases where the assumed values for LL coefficients agree with the results of the free fits, the NLL coefficients can be obtained with much smaller errors.

So far there exist no analytic predictions for the separate regions (i) and (ii). Therefore we can compare only the total contributions with the leading logarithm results of [8]. These are for the $C_F^2$ coefficients: $R_{24} = 2$, $R_{23} = 6$ and $R_{22} = (13/2 - 2\pi^2) = -13.24$ and for the $C_F N_C$ terms: $R_{24} = 0$, $R_{23} = 11/3$ and $R_{22} = -(169/36 - \pi^2/3) = -1.405$. As we see from Tab. 6, only the LL coefficients are consistent with our results $R_{24} = 1.985 \pm 0.004$ for the $C_F^2$ coefficient, and $R_{24} = 0.140 \pm 0.009$ for the $C_F N_C$ coefficient, respectively. The NLL contributions deviate considerably. The $C_F N_C$ coefficient $R_{23}$ is determined in region (ii) only with a very large error. This is the reason for the deviation of $R_{23}$ in the sum. If we interpret the result of $R_{23}$ in region (ii) as $R_{23} = 0$, the NLL coefficient from the singular region is very well determined and agrees with $R_{23} = 11/3$ as predicted. The NLL terms in the sum are determined with the leading logarithm hypothesis $R_{24} = 2$ and $R_{24} = 0$ (for the $C_F^2$ and $C_F N_C$ coefficients, respectively). In addition we have calculated the NNLL coefficients $R_{22}$ and $R_{21}$ with the leading logarithm constraints for $R_{24}$ and $R_{23}$. The results are different now and still with large errors. $R_{22}(C_F N_C)$ agrees now better with the leading logarithm result.

If one compares the leading logarithm predictions of $R_{24}$ and $R_{23}$ for different event shape distributions, as for example thrust [8] and heavy jet mass [9], one notices that they agree [8]. They differ, however, in $R_{22}$. All these leading logarithm derivations, including the derivation for the 2-jet cross section in [6], start from the jet mass distribution which was calculated to NLL accuracy in [20]. Therefore the coefficients $R_{24}$ and $R_{23}$ must agree in these calculations. The LL and NLL coefficients of the 2-jet cross sections have been calculated earlier in [19] with completely different methods and with the same result. We notice that also $R_{22}$ for $C_F N_C$ is universal for the thrust and heavy jet mass distributions and the 2-jet cross section, namely $R_{22} = (\pi^2/3 - 169/36)$.

For calculating the LL and NLL coefficients for the energy-energy correlation $d\Sigma/d\cos\chi$ we also use the differential distribution in $\cos\chi$ where $\chi$ is the angle between two jets. This has the form (15), i.e.

$$\frac{1}{\sigma_0}\frac{d\Sigma}{d\cos\chi} = \frac{\alpha_s}{2\pi}\frac{dR_1}{d\cos\chi} + \left(\frac{\alpha_s^2}{2\pi}\right)^2 \frac{dR_2}{d\cos\chi} \tag{17}$$

and the $O(\alpha_s^2)$ term has, similar to (16), the expansion

$$\frac{dR_2}{d\cos\chi} = \sum_{m=1}^{4} mR_{2m}\frac{1}{\eta}L_\eta^{m-1} + \frac{dD_2}{d\eta} \tag{18}$$

where $\eta = (1+\cos\chi)/2$ and $L_\eta = \ln(1/\eta)$. Note that in this case we define $L_\eta$ with a different sign as compared with $L_\tau$ such that $L_\eta$ is positive. The LL, NLL and NNLL coefficients $R_{24}$, $R_{23}$ and $R_{22}$ for the $C_F^2$ and $C_FN_C$ terms have been calculated by many authors in the past [11, 12] using the leading logarithm formalism. In the complete $O(\alpha_s^2)$ theory they have been computed by Ellis et al. [12] some time ago and recently again by Ellis and Clay [21] with much higher accuracy. In the older calculation Ellis et al. claimed to have found clear disagreement in the $C_F^2$ part of $R_{22}$, whereas all other coefficients agreed with the leading logarithm predictions within their numerical accuracy. In the more recent work [21] the authors found good agreement between leading logarithm predictions and their results for both the $C_F^2$ and the $C_FN_C$ part from fitting their numerical results to the expansion (18). In both the recent and the older work the authors used the subtraction method with the matrix elements of [13].

In our work we introduce a slicing parameter $y_0 = 0.1$ as in the calculation of the thrust distribution. With this large value for $y_0$ the contribution which comes from region (iii) is negligible and has no effect on the determination of the coefficients $R_{2m}$ in (18). We calculate the contributions from the two regions (i) and (ii), where (i) and (ii) denote the contributions of the singular and non-singular terms with $y_{kl}$ below the cut $y_0$. All NLL coefficients are obtained with specific assumptions for the values of the corresponding LL coefficients.

The results are collected in Tab. 7. The first and second row give the coefficients $R_{2m}$, $(m = 1,2,3,4)$ for the singular and non-singular terms. The results for the singular terms are rather stable with small errors for the coefficients. They can be compared with results from a completely analytic calculation of the singular contributions in [22] without the $y_0$ cut, i.e. over the full 4-parton phase space. The results for the $C_F^2$ terms from [22] were: $4R_{24} = -1/3$, $3R_{23} = 3/2$, $2R_{22} = -8.456$, $R_{21} = 14.34$ and for the $C_FN_C$ terms: $4R_{24} = -1/12$, $3R_{23} = 11/12$, $2R_{21} = 3.204$ and $R_{21} = -8.269$. Comparing this with our fit results, we see that we achieved good agreement for both the $C_F^2$ and the $C_FN_C$ terms including $R_{22}$. Only for $R_{21}$ we observe deviations up to 30%. This had to be expected since the coefficients have alternating signs so that a precise determination of the last coefficient is quite difficult. These results can also be considered as a test of our method, in particular with respect to introducing the slicing cut-off which obviously cuts out only an unessential part of the 4-parton phase space.

There are no analytic results for region (ii). So we can compare only the sum of both contributions. The LL coefficients agree very well with $4R_{24} = -1/2$ ($C_F^2$ part) and $R_{24} = 0$ ($C_FN_C$ part) within the error which is 27% for the $C_F^2$ coefficient. The NLL and NNLL coefficients show no resemblance to the leading logarithm prediction which are $3R_{23} = 9/4$, $2R_{22} = -(17/4 + \pi^2/6) = -5.859$ and $3R_{23} = 11/12$, $2R_{22} = (35/72 - \pi^2/12) = -0.336$ for the two colour factors, respectively, although the LL values for $R_{24}$ were assumed as constraint. Since the NLL and NNLL coefficients have opposite signs they compensate each other and can not be determined accurately enough. We also tried several other hypotheses for the LL coefficients. For example, if we assume $R_{24} = -7/12$ for the leading $C_F^2$ coefficient, we obtain: $3R_{23} = 2.478 \pm 0.140$, $2R_{22} = -17.44 \pm 2.14$ and $R_{21} = 37.4 \pm 8.6$. In this case the NLL and NNLL coefficients change sign as compared to the results in Tab. 7. This shows that the NLL terms can be determined only when the LL coefficient is known exactly. From this exercise we can not exclude that the leading logarithm result $4R_{24} = -1/2$ does perhaps not agree with the LL coefficient of a complete calculation. In addition in the last rows we made the LL hypothesis for $R_{24}$ and $R_{23}$ and determined $R_{22}$ and $R_{21}$. Now at least the sign of $R_{22}$ comes out correctly for both colour factors.

We have to keep in mind, however, that with our method we calculate the energy-energy correlation on the basis of three particles in the final state, and not four, as it would be appropriate for a 4-parton final state. This reduction to a 3-jet final state is unavoidable in some part of the 4-parton phase space. Otherwise the cancellation of virtual and real soft and collinear divergences is not possible. Therefore the question remains whether our results are sensitive to the choice of the slicing parameter $y_0$. In this respect two situations must be distinguished. If $y_0$ is above the region where the leading terms in $L_\eta$ have been determined (this is the procedure followed here) there exists no dependence on $y_0$ when the value of this parameter is large enough as compared to the fitting region. We checked explicitly that all coefficients change only by amounts smaller than the quoted errors when $y_0$ is reduced to $y_0 = 10^{-2}$. When $y_0$ is chosen small and inside the fitting region for the leading logarithm expansion, then the discontinuity due to switching from 3- to 4-parton final states prevents obtaining reasonable fits without taking account of non-logarithmic terms. Only for the leading coefficients consistent results can be obtained in this case. To choose $y_0$ below and outside the fitting region is impossible for the energy-energy correlation, or, as in the case of the thrust distribution, would result in too large errors.

So far we considered only the $C_F^2$ and $C_F N_C$ contributions in the colour decomposition (10). The coefficients of the $C_F N_f$ contribution have been determined in the same way. These contributions come from the diagonal part of the $q\bar{q}q\bar{q}$ final state. They have a simpler singularity structure. Only single-pole terms appear and partial fractioning is not needed. Also the non-singular contributions from region (ii) can be treated analytically and therefore have been combined with the singular contributions from region (i) (see [1] for further details). The less singular behaviour of these terms has the consequence that the LL coefficients $R_{24}$ vanish for this colour factor. We have calculated the coefficients $R_{23}, R_{22}$ and $R_{21}$ for the thrust distribution and the energy-energy correlation and in addition $R_{20}$ for the jet cross sections.

The results for the coefficients of the thrust distribution are collected also in Tab. 6 together with the results of the other colour factors. They agree reasonably well with the LL prediction $R_{23} = -2/3$ and $R_{22} = 11/18$ [8]. If $3R_{23} = -2$ is put in as a constraint we obtain $2R_{22} = 1.239 \pm 0.018$ and $R_{21} = 2.90 \pm 0.15$, which is now in very good agreement with $2R_{22} = 11/9$ and $R_{21}$ is close to the corresponding coefficient $R_{21} = 5/2$ in the two-jet cross section [19].

The coefficients for the energy-energy correlation are presented in Tab. 7. The LL results are $R_{23} = -1/18$ and $R_{22} = -1/72$ [12, 22]. The value for $R_{23}$ is confirmed by our fit. With the constraint $3R_{23} = -1/6$ we obtain $2R_{22} = -0.0102 \pm 0.0077$ which agrees now approximately with the prediction $2R_{22} = -1/36$. The prediction for the next coefficient is $R_{21} = (11/24 - \pi^2/18) = -0.090$ [22]. In our fit we obtain at least the same sign and also a small value.

The results for the jet rates with the different algorithms are collected separately in Tab. 8. For every jet algorithm and recombination scheme we have two lines. In the first line all coefficients $R_{2i}, i = 0, 1, 2, 3$ are determined from the same fit. In the second line $R_{23}$ is constrained by its LL value as given in the table. Let us first look at the results for the 3-jet cross sections. Here in all three cases, JADE-E0, JADE-E and Durham-E0, we find very good agreement with the LL results, which are $R_{23} = 7/9, 7/9, 1/3$, respectively. Unfortunately there exist no predictions for the next-to-leading coefficients. For the 4-jet cross section we find similar results, i.e. good agreement with $R_{23} = -1/9$ for both the JADE and Durham algorithm (for $n = 4$ the results for the E0 and E schemes must coincide). Also for $n = 3$ we obtained the same coefficients in the JADE-E0 and E scheme. One further check is possible if we calculate the coefficients of the 2-jet cross section which are equal to the negative sum of the 3- and 4-jet coefficients. In the JADE algorithm they are $R_{23} = -0.628 \pm 0.007$ and $R_{22} = 0.64 \pm 0.12$ (with $R_{23}$ constrained). This agrees with results from analytic calculations of the 2-jet cross section in the JADE algorithm [19]: $R_{23} = -2/3$ and $R_{22} = 11/18$. In total we conclude that for the $C_F N_f$ coefficients the agreement with the leading logarithm results or pure analytic calculations is quite good.

# 4 Concluding Remarks

In the last section we have seen that our results for the LL and NLL coefficients have different accuracy depending on the kind of jet cross section, event shape distribution or colour factor. Therefore conclusions concerning the verification of leading logarithm predictions have also different quality. The clearest statement can be made for the $C_F N_f$ coefficients. Here we found perfect agreement with the leading logarithm predictions for all studied jet cross sections and event shape distributions. For the other colour factors the main interest concerns those jet cross sections and shape distributions which can be summed in the leading logarithm approximation, i.e. the jet cross sections in the Durham scheme, the thrust distribution and the energy-energy correlation.

The LL, NLL and NNLL coefficients of the Durham cross sections were obtained with reasonably small errors. Both the LL and NLL coefficients agreed perfectly with the leading logarithm predictions (see Tabs. 3 and 5). The main contributions came from the singular region (i); but also the other regions contributed. It is known that the Durham algorithm by using the transverse momentum as jet resolution is adapted best to the underlying infrared and collinear singularity structure of the tree diagram contributions.

For the thrust distribution, although the accuracy was similar as for the Durham cross sections, we could confirm only the LL coefficient. The NLL coefficients show appreciable deviations from the leading logarithm predictions. On the basis of our results, however, it would be premature to claim that deviations in the NLL coefficients of the thrust distribution have been established.

For the energy-energy correlation we have compared our results in the singular region with an independent analytic calculation [22]. We found good agreement up to the NNLL coefficients. Concerning the complete coefficients from all regions the situation is similar as for the thrust distribution. Only the LL coefficients agree with the corresponding leading logarithm prediction.

The jet cross sections for the JADE algorithm can not be resummed. This has been confirmed by our results for the $C_F^2$ LL coefficient of the 3-jet cross section. We found that this coefficient does not agree with the leading logarithm result as reported by Catani [5]. The coefficients for the 4-jet cross section do agree, however.

## Table Captions

**Tab. 1:** Leading logarithm coefficients for 3- and 4-jet rates in the JADE-E0 scheme.

**Tab. 2:** Leading logarithm coefficients for 3- and 4-jet rates in the JADE-E scheme.

**Tab. 2:** Leading logarithm coefficients for 3- and 4-jet rates in the Durham-E0 scheme.

**Tab. 4:** Contributions to the leading logarithm coefficients for the 3-jet rate in the JADE-E0 scheme.

**Tab. 5:** Contributions to the leading logarithm coefficients for the 3-jet rate in the Durham-E0 scheme.

**Tab. 6:** Contributions to the leading logarithm coefficients for the thrust distribution.

**Tab. 7:** Contributions to the leading logarithm coefficients for the energy-energy correlation.

**Tab. 8:** Leading logarithm coefficients for the $C_F N_f$ contribution to the 3- and 4-jet rates.

Table 1:

| JADE Algorithm–E0 Scheme | | | | | | |
|---|---|---|---|---|---|---|
| $n$ | $colour factor$ | $R_{24}$ | $R_{23}$ | $R_{22}$ | $R_{21}$ | $R_{20}$ |
| 3 | $C_F^2$ | $-1.63 \pm 0.69$ | $-26.95 \pm 0.90$ | | | |
| | $C_F N_C$ | $-2.6 \pm 1.9$ | $16.6 \pm 2.5$ | | | |
| 4 | $C_F^2$ | $1.46 \pm 0.15$ | $6.17 \pm 0.19$ | $-3.1 \pm 0.25$ | $-37.0 \pm 3.3$ | $32.0 \pm 10.0$ |
| | $C_F N_C$ | $0.165 \pm 0.052$ | $1.57 \pm 0.07$ | $7.64 \pm 0.09$ | $21.7 \pm 1.2$ | $29.4 \pm 3.6$ |

Table 2:

| JADE Algorithm–E Scheme | | | | | | |
|---|---|---|---|---|---|---|
| $n$ | $colour factor$ | $R_{24}$ | $R_{23}$ | $R_{22}$ | $R_{21}$ | $R_{20}$ |
| 3 | $C_F^2$ | $-2.79 \pm 0.27$ | $17.8 \pm 2.1$ | | | |
| | $C_F N_C$ | $-1.6 \pm 2.0$ | $7.3 \pm 2.6$ | | | |
| 4 | $C_F^2$ | $1.46 \pm 0.15$ | $6.17 \pm 0.19$ | $-3.1 \pm 0.25$ | $-37.0 \pm 3.3$ | $32.0 \pm 10.0$ |
| | $C_F N_C$ | $0.165 \pm 0.052$ | $1.57 \pm 0.07$ | $7.64 \pm 0.09$ | $21.7 \pm 1.2$ | $29.4 \pm 3.6$ |

Table 3:

| DURHAM Algorithm–E0 Scheme | | | | | | |
|---|---|---|---|---|---|---|
| $n$ | $colour factor$ | $R_{24}$ | $R_{23}$ | $R_{22}$ | $R_{21}$ | $R_{20}$ |
| 3 | $C_F^2$ | $-1.29 \pm 0.46$ | $-6.34 \pm 0.60$ | $-9.48 \pm 0.78$ | $21.8 \pm 11.2$ | $7.9 \pm 39.0$ |
| | $C_F N_C$ | $-0.32 \pm 0.25$ | $-3.19 \pm 0.33$ | $-6.57 \pm 0.42$ | $-11.3 \pm 6.0$ | $8.9 \pm 21.0$ |
| 4 | $C_F^2$ | $0.459 \pm 0.045$ | $3.108 \pm 0.058$ | $3.82 \pm 0.08$ | $-13.4 \pm 0.99$ | $36.7 \pm 3.1$ |
| | $C_F N_C$ | $0.062 \pm 0.011$ | $1.097 \pm 0.014$ | $7.368 \pm 0.019$ | $25.70 \pm 0.25$ | $36.12 \pm 0.78$ |

Table 4:

| JADE Algorithm–E0 Scheme | | | | | |
|---|---|---|---|---|---|
| $region$ | $colour factor$ | $R_{24}$ | $R_{23}$ | $R_{24}$ | $R_{23}$ ($R_{22}$) |
| $(i)$ | $C_F^2$ | $-3.974 \pm 0.006$ | $-11.25 \pm 0.18$ | $-4.0$ | $-12.084 \pm 0.013$ |
| | $C_F N_C$ | $-0.175 \pm 0.002$ | $-5.593 \pm 0.061$ | $-1/6$ | $-5.254 \pm 0.004$ |
| $f(x_1, x_2)$ | $C_F^2$ | $0.0020 \pm 0.0002$ | $0.0664 \pm 0.0071$ | $0.0$ | $0.0$ ($-3.971 \pm 0.001$) |
| | $C_F N_C$ | $-0.499 \pm 0.001$ | $0.019 \pm 0.042$ | $-1/2$ | $0.0$ ($0.308 \pm 0.007$) |
| $(ii)$ | $C_F^2$ | $0.0234 \pm 0.0029$ | $0.638 \pm 0.096$ | $0.0$ | $-0.1347 \pm 0.0066$ |
| | $C_F N_C$ | $-1.506 \pm 0.055$ | $-46.2 \pm 1.5$ | $-3/2$ | $-46.02 \pm 0.08$ |
| $(iii)$ | $C_F^2$ | $0.515 \pm 0.017$ | $0.62 \pm 0.56$ | $1/2$ | $0.121 \pm 0.039$ |
| | $C_F N_C$ | $-0.2237 \pm 0.0089$ | $-0.41 \pm 0.29$ | $-1/6$ | $1.441 \pm 0.021$ |

Table 5:

| DURHAM Algorithm–E0 Scheme | | | | | |
|---|---|---|---|---|---|
| *region* | *colourfactor* | $R_{24}$ | $R_{23}$ | $R_{24}$ | $R_{23}$ ($R_{22}$) |
| $(i)$ | $C_F^2$ | $-1.997 \pm 0.008$ | $-9.03 \pm 0.26$ | $-2.0$ | $-9.124 \pm 0.019$ |
| | $C_F N_C$ | $-0.0816 \pm 0.0031$ | $-2.78 \pm 0.10$ | $-1/12$ | $-2.838 \pm 0.007$ |
| $f(x_1, x_2)$ | $C_F^2$ | $0.00075 \pm 0.00035$ | $0.0098 \pm 0.0012$ | $0.0$ | $0.0$ ($-4.209 \pm 0.002$) |
| | $C_F N_C$ | $-0.0832 \pm 0.0009$ | $0.012 \pm 0.028$ | $-1/12$ | $0.0$ ($1.887 \pm 0.005$) |
| $(ii)$ | $C_F^2$ | $0.008 \pm 0.001$ | $0.324 \pm 0.042$ | $0.0$ | $0.0464 \pm 0.0030$ |
| | $C_F N_C$ | $0.0110 \pm 0.0088$ | $0.396 \pm 0.260$ | $0.0$ | $0.070 \pm 0.017$ |
| $(iii)$ | $C_F^2$ | $1.005 \pm 0.018$ | $3.17 \pm 0.60$ | $1.0$ | $3.012 \pm 0.042$ |
| | $C_F N_C$ | $-0.033 \pm 0.010$ | $-0.52 \pm 0.32$ | $0.0$ | $0.518 \pm 0.024$ |

Table 6:

| Coefficients for Thrust Distribution | | | | | |
|---|---|---|---|---|---|
| *region* | *colourfactor* | $4R_{24}$ | $3R_{23}$ | $2R_{22}$ | $R_{21}$ |
| $(i)$ | $C_F^2$ | $5.332 \pm 0.009$ | $12.21 \pm 0.24$ | $18.7 \pm 2.3$ | $23.6 \pm 6.9$ |
| | $C_F N_C$ | $1.993 \pm 0.006$ | $10.80 \pm 0.17$ | $-25.6 \pm 1.6$ | $-25.6 \pm 4.9$ |
| $(ii)$ | $C_F^2$ | $2.610 \pm 0.015$ | $-0.90 \pm 0.43$ | $-10.7 \pm 1.7$ | $-27.0 \pm 12.0$ |
| | $C_F N_C$ | $-1.73 \pm 0.18$ | $1.9 \pm 3.7$ | $-13.5 \pm 23.6$ | $46.0 \pm 48.0$ |
| *sum* | $C_F^2$ | $7.941 \pm 0.018$ | $12.75 \pm 0.02$ | $18.93 \pm 0.41$ | $36.3 \pm 1.8$ |
| | $C_F N_C$ | $0.560 \pm 0.038$ | $3.15 \pm 0.37$ | $-77.6 \pm 3.6$ | $-153.7 \pm 8.8$ |
| | $C_F N_f$ | $0.0$ | $-1.9900 \pm 0.0051$ | $1.53 \pm 0.17$ | $4.02 \pm 0.65$ |
| *sum* | $C_F^2$ | $8.0$ | $18.0$ | $107.6 \pm 6.4$ | $405.0 \pm 52.0$ |
| | $C_F N_C$ | $0.0$ | $11.0$ | $-2.2 \pm 1.2$ | $28.5 \pm 5.9$ |
| | $C_F N_f$ | $0.0$ | $-2.0$ | $1.239 \pm 0.018$ | $2.90 \pm 0.15$ |

Table 7:

| Coefficients for Energy-Energy Correlation | | | | | |
|---|---|---|---|---|---|
| *region* | *colourfactor* | $4R_{24}$ | $3R_{23}$ | $2R_{22}$ | $R_{21}$ |
| $(i)$ | $C_F^2$ | $-0.3304 \pm 0.0042$ | $1.5066 \pm 0.0080$ | $-8.37 \pm 0.15$ | $11.28 \pm 0.74$ |
| | $C_F N_C$ | $-0.078 \pm 0.013$ | $0.946 \pm 0.017$ | $2.76 \pm 0.30$ | $-5.72 \pm 1.38$ |
| $(ii)$ | $C_F^2$ | $-0.13 \pm 0.10$ | $-1.634 \pm 0.024$ | $16.56 \pm 0.30$ | $-55.2 \pm 1.4$ |
| | $C_F N_C$ | $0.0312 \pm 0.0052$ | $-1.042 \pm 0.092$ | $9.3 \pm 1.6$ | $-27.4 \pm 7.4$ |
| *sum* | $C_F^2$ | $-0.46 \pm 0.13$ | $-0.220 \pm 0.028$ | $9.60 \pm 0.50$ | $-49.6 \pm 2.4$ |
| | $C_F N_C$ | $0.0 \pm 0.015$ | $-0.090 \pm 0.018$ | $11.90 \pm 0.32$ | $-32.6 \pm 1.4$ |
| | $C_F N_f$ | $0.0$ | $-0.169 \pm 0.006$ | $0.037 \pm 0.022$ | $-0.54 \pm 0.11$ |
| *sum* | $C_F^2$ | $-1/2$ | $9/4$ | $-34.2 \pm 3.4$ | $139. \pm 26.$ |
| | $C_F N_C$ | $0.0$ | $11/12$ | $-5.2 \pm 1.2$ | $38.6 \pm 9.2$ |
| | $C_F N_f$ | $0.0$ | $-1/6$ | $-0.0102 \pm 0.0077$ | $-0.330 \pm 0.019$ |

Table 8:

| $C_F N_f$ Coefficients for Jet Rates | | | | | |
|---|---|---|---|---|---|
| $n$ | *algorithm* | $R_{23}$ | $R_{22}$ | $R_{21}$ | $R_{20}$ |
| 3 | JADE E0 | $0.747 \pm 0.006$<br>$7/9$ | $-1.11 \pm 0.13$<br>$-0.14 \pm 0.09$ | $-12.1 \pm 1.3$<br>$-2.50 \pm 0.08$ | $-31.2 \pm 4.4$<br>$-0.36 \pm 0.40$ |
| 3 | JADE E | $0.740 \pm 0.006$<br>$7/9$ | $-1.34 \pm 0.13$<br>$-0.14 \pm 0.12$ | $-14.4 \pm 1.4$<br>$-2.49 \pm 0.08$ | $-38.2 \pm 4.4$<br>$-0.37 \pm 0.40$ |
| 3 | DURHAM E0 | $0.36 \pm 0.11$<br>$1/3$ | $2.0 \pm 3.7$<br>$1.04 \pm 0.21$ | $13.5 \pm 42.0$<br>$3.9 \pm 4.9$ | $37.0 \pm 160.$<br>$5.8 \pm 26.5$ |
| 4 | JADE | $-0.1187 \pm 0.0036$<br>$-1/9$ | $-0.70 \pm 0.12$<br>$-0.5016 \pm 0.0031$ | $-2.3 \pm 1.2$<br>$-0.26 \pm 0.07$ | $-6.3 \pm 4.0$<br>$0.70 \pm 0.38$ |
| 4 | DURHAM | $-0.1108 \pm 0.0019$<br>$-1/9$ | $-1.07 \pm 0.06$<br>$-1.0878 \pm 0.0018$ | $-3.6 \pm 0.7$<br>$-3.667 \pm 0.039$ | $-3.1 \pm 2.1$<br>$-3.6 \pm 0.20$ |